text of paper:
 \documentstyle[preprint,eqsecnum,aps]{revtex}
 
\begin{document}
 \baselineskip=0.6 cm
 \title{SPIN-LADDERS WITH SPIN GAPS: A DESCRIPTION OF A CLASS OF
 CUPRATES}
 \author{Sudha Gopalan, T. M. Rice, and M. Sigrist\cite{current}}
 \address{Theoretische Physik, Eidgen\"ossische Technische
 Hochschule - H\"onggerberg,\\ CH-8093 Z\"urich, Switzerland.
\\(Received 17 November 1993.)}
\maketitle
 \begin{abstract}
 \noindent

We investigate the magnetic properties of the Cu-O planes
in stoichiometric
Sr$_{n-1}$Cu$_{n+1}$O$_{2n}$ (n=3,5,7,...) which consist
of CuO double chains periodically
intergrown within the CuO$_2$ planes. The double chains
break up the two-dimensional antiferromagnetic planes
into Heisenberg spin ladders with
$n_r=\frac{1}{2}(n-1)$ rungs and  $n_l=\frac{1}{2}(n+1)$
legs and described by the usual antiferromagnetic
coupling J inside each ladder and a
weak and frustrated interladder coupling J$^\prime$.
The resulting lattice is a new two-dimensional trellis
lattice. We first examine the spin excitation spectra of
isolated quasi one dimensional Heisenberg ladders
which exhibit a gapless spectra when $n_r$ is even
and $n_l$ is odd ( corresponding to n=5,9,...) and
a gapped spectra when $n_r$ is odd and $n_l$ is even
(corresponding to n=3,7,...).
We use the bond operator representation of
quantum $S=\frac{1}{2}$ spins in a mean field treatment
with self-energy corrections  and obtain a spin gap of
$\approx \frac{1}{2} J$ for the simplest single
rung ladder (n=3), in agreement with numerical estimates.
We also present results of the dynamical
structure factor S(q,$\omega$).  The spin gap decreases
considerably on increasing the width of the ladders.
For a double ladder with four legs and three rungs (n=7)
we obtain a spin gap of only 0.1J. However, a frustrated
coupling, such as that of a trellis lattice,
introduced between the double ladders
leads to an enhancement of the gap. Thus stoichiometric
Sr$_{n-1}$Cu$_{n+1}$O$_{2n}$ compounds with n=3,7,11,...
will be frustrated quantum antiferromagnets with a
quantum disordered or spin-liquid ground state.
 \end{abstract}

 \section{INTRODUCTION}

The so-called "infinite layer" or "all layer" compound
SrCuO$_2$ represents a new family of the Cu-O
superconductors. This compound crystallizing in the
tetragonal structure is characterized by an infinite
stacking of CuO$_2$ planes with intervening Sr layers
without oxygen representing the simplest possible charge
reservoir. The infinite layer SrCuO$_2$ \cite{siegrist}
are synthesized under extreme conditions involving high
temperatures and high oxygen pressures and superconductivity
in the Cu-O layers is induced by modification of the
intervening Sr layers. The Cu-O layers can be either
hole or electron doped depending on the dopants introduced
in the Sr layer. The high pressure forms of SrCuO$_2$ are
however unstable above certain temperatures and a homologous
series of oxides formulated as Sr$_{n-1}$Cu$_{n+1}$O$_{2n}$
(with n=3,5,7,9,.....) begins to be
mixed in with the parent (n=$\infty$) phase.

The homologous series of Cu-rich high pressure phases
Sr$_{n-1}$Cu$_{n+1}$O$_{2n}$ were recently studied
\cite{takano} and were shown to consist of
parallel lines of CuO-double chains periodically
intergrown within the CuO$_2$ sheets as illustrated in
Fig. 1(a). The occurrence of these double chains was
interpreted as resulting from a periodic shear operation
with a shear vector of $\frac{1}{2}\langle$110$\rangle$
in the parent (n=$\infty$) phase. This can also be
visualised as the appearance of domain walls within
the planes. The structure of one such sheet for a
general n is shown in  Fig. 1(a). The Cu-atoms are shown
as big black dots while the oxygen atoms are located
at all the points of intersections of the straight
lines. The Sr atoms (not shown in the figure) are
located at the centers of the squares which are empty
and in planes displaced by $\pm$ c/2 with respect to
the copper-oxide sheet shown with c being the lattice
constant perpendicular to the Cu-O planes. The dotted
circles are the square co-ordinated Cu-atoms (like in
the parent compound) and there are respectively 0,1,2,...
such Cu-atoms in between two double chains for n=3,5,7...
Thus the introduction of the CuO double chains
periodically in the matrix of square planar co-ordinated
CuO$_2$ produces a superlattice geometry with unit cell
parameters (na$\times$a).

The double chains in Sr$_{n-1}$Cu$_{n+1}$O$_{2n}$
compounds affect dramatically the magnetic properties
of the copper-oxide planes. In the stoichiometric
compounds the Cu sites are singly occupied (in the hole
notation) and the O-sites are empty. This can be
modelled by a S=1/2 Heisenberg model on the new lattice,
the {\it trellis lattice}, formed from the Cu-sites. The
exchange interaction between Cu-atoms which are both not
located along the double chains is given by that of the
bulk (n=$\infty$) value of J. Along the double chains
however, two Cu-ions (such as A and B of Fig. 1(a)) are
connected via an O-site by 90$^\circ$ bonds which gives
rise to a ferromagnetic exchange \cite{anderson}.
The superexchange path between these Cu-ions is through
two orthogonal O-orbitals and this introduces
a Hund's rule contribution leading to a ferromagnetic
exchange J$^\prime$ ($<$ 0) given in perturbation theory by
 \begin{equation}
  J^\prime  = \frac{8t^4_{pd}}{{\Delta}^2}[ \frac{1}{E_T
+ 2\Delta} -\frac{1}{E_S + 2 \Delta} ]
 \end{equation}
where t$_{pd}$ is the nearest neighbor hopping in the
plane between the Cu3d(x$^2$-y$^2$) to the O2p(x,y) orbitals,
$\Delta$ = $\epsilon_p -\epsilon_d$ with $\epsilon_p$ and
$\epsilon_d$ being respectively the on-site energies of
the O2p(x,y) and the Cu3d(x$^2$-y$^2$) levels, and
E$_T$ and E$_S$ are respectively the triplet and
singlet levels. Using the values of E$_T$ = 1.8eV and
E$_S$ = 7.3eV obtained from the level splittings
\cite{atomic} assuming an onsite Coulomb repulsion of
U$_p$ =4eV \cite{hybertsen} and taking the standard
values of $t_{pd}$=1.3eV and $\Delta$ = 3.3eV
\cite{hybertsen} we obtain $J^\prime$/J=0.1-0.2.

Thus the planes in Sr$_{n-1}$Cu$_{n+1}$O$_{2n}$
are broken up into Heisenberg ladders with a weak
and frustrated interladder coupling as illustrated
in Fig. 1(b). There are 0,1,2... vertical chains
(separated by the width a) cutting
the dashed region in Fig. 1(b) corresponding to
n=3,5,7... . The lattice that results is a trellis lattice
consisting of individual ladders with $n_r=\frac{1}{2}(n-1)$
rungs and $n_l=\frac{1}{2}(n+1)$ legs and
coupled to each other through zig-zag couplings J$^\prime$.
Since J$^\prime \ll$J we can to a first approximation
neglect J$^\prime$ and divide the Cu-O
planes into independent sets of ladders with $n_r$=1,2,3...
rungs and n$_l$=2,3,4.....legs corresponding to n=3,5,7... .
The compounds with odd n$_r$ and even n$_l$
(corresponding to n=3,7,11,...) and those with even
n$_r$ and odd n$_l$(corresponding to n=5,9,13,...) exhibit
different spin excitation spectra, the
former having a spin gap with a shortrange exponentially
decaying magnetic correlation function
while the latter are gapless with a longrange
powerlaw decay of the correlation function
(like in the case of a single chain corresponding to n=1).

It was argued in ref. \cite{rice} that
Sr$_{n-1}$Cu$_{n+1}$O$_{2n}$ compounds
with n=3,7,11... would be frustrated quantum
antiferromagnets and spin liquids. This was
motivated by the numerical studies \cite{hirsch},\cite{dag}
on isolated Heisenberg single-rung antiferromagnetic ladders
with couplings J which exhibited a gap of $\approx \frac{1}{2}
J$. Here we will follow an alternate treatment in terms of
bond operators formulated \cite{sach} to study
two-dimensional dimerised Heisenberg systems. In section II
we will emphasize some of the important features of
these bond operaters and use them in a mean-field
treatment to study the simplest single rung
Heisenberg ladder with two legs.
In section III we report the results
of the calculations on the spin excitation spectrum,
spin gap, ground state energy and the dynamical structure
factor $S(k,\omega)$. We extend
the calculations to double ladders and also to
periodic ladders in section IV. We show that the spin gap
decreases considerably on increasing the width (rungs) of
the ladders. In section V we study the effect of
frustration on two spin-ladders and show that any
non-zero frustrated coupling leads to an
enhancement of the gap. This important result points
towards the stabilization of a spin-liquid ground
state in a trellis lattice. Such lattices will be
realizations of short range RVB (Resonating Valence
Bond) ground states in a S=$\frac{1}{2}$ system
\cite{anderson2}.

 \section{SINGLE SPIN-LADDER}

We investigate here the properties of a single spin-
$\frac{1}{2}$ ladder, shown in Fig. 2(a) which is
described by the standard Heisenberg antiferromagnetic model
 \begin{equation}
 H = J \sum_{i} {\bf S}_{l_i }\cdot {\bf S}_{r_i} +
 \lambda J \sum_{i,m=l,r} {\bf S}_{m_i }\cdot
{\bf S}_{m_{i+1}} ,
 \end{equation}
where we take J to be the strength of the interaction
along the rungs(i) of the ladder and $\lambda$J the
interaction along the legs of the ladder.  ${\bf
 S}_{l_i} $ and ${\bf S}_{r_i }$ are
respectively spin-$\frac{1}{2}$ operators at
the left(l) and right(r) -hand sites on each rung i
of the ladder. In the limit of $\lambda$=0 (strong-
coupling limit) the Hamiltonian of Eq.(2.1)
reduces to a sum over contributions from independent
two spin rungs. Thus it would be natural to tackle
this problem from the limit of singlet dimers
placed on the rungs and then switch on the interaction
between them. It will be seen later that this starting
point leads to reasonable results even for the case of
$\lambda$=1 which is of interest here. We follow the
bond-operator representation of quantum S=$\frac{1}{2}$ -
spins introduced \cite{sach} to study specifically the
properties of dimerized phases. We emphasize here
the essential features of the bond operators.

We consider two  S = $\frac{1}{2}$ spins ${\bf S}_l$
and ${\bf S}_r$ placed on each rung. The
Hilbert space is spanned by four states which can be
combined to form the singlet $|s>$ and the three triplet
$|t_x>$, $|t_y>$, and $|t_z>$ -states defined as being created
out of the vacuum $|0>$ by the singlet and triplet
creation operators
 \begin{eqnarray}
 |s> &=& s^\dagger |0> =  \frac{1}{\sqrt{2}}
(| \uparrow\downarrow > - |\downarrow\uparrow >), \nonumber
 \\
 |t_x>&=& t^\dagger_x |0>=-\frac{1}{\sqrt{2}}
(| \uparrow\uparrow > - |\downarrow\downarrow >), \nonumber
 \\
 |t_y>&=& t^\dagger_y |0>=\frac{{\it i}}{\sqrt{2}}
(|\uparrow\uparrow > + |\downarrow\downarrow >),
 \\
 |t_z>&=& t^\dagger_z |0>=\frac{1}{\sqrt{2}}
(| \uparrow\downarrow > + |\downarrow\uparrow >). \nonumber
 \end{eqnarray}

 A representation of the spins ${\bf S}_l$ and ${\bf S}_r$
 in terms of these singlet and triplet operators is given by,
 \begin{eqnarray}
 S_{l\alpha}&=& \frac{1}{2} (s^\dagger t_\alpha
+ t^\dagger_\alpha s
 - {\it i} {\epsilon}_{\alpha\beta\gamma} t^\dagger_\beta t_\gamma),
 \\
 S_{r\alpha}&=& \frac{1}{2} (-s^\dagger t_\alpha
- t^\dagger_\alpha s
 - {\it i} {\epsilon}_{\alpha\beta\gamma} t^\dagger_\beta t_\gamma),
 \end{eqnarray}
 where $\alpha$, $\beta$, and $\gamma$ represent
respectively the components along the x,y, and z-axes
and $\epsilon$ is the Levi-Civita symbol
 representing the totally antisymmetric tensor.
Henceforth, it is assumed that all repeated indices over
$\alpha$, $\beta$, and $\gamma$ are summed over.

A constraint of the form,
 \begin{equation}
 s^\dagger s + t^\dagger_\alpha  t_\alpha =  1,
 \end{equation}
 is introduced for each dimer in order to restrict the
 physical states to either singlets or triplets.
Taking the singlet and triplet operators at
each site to satisfy the bosonic commutation relations,
 \begin{equation}
 [s , s^\dagger] = 1   , \;  [t_{\alpha} ,
  t^\dagger_{\beta}] = \delta _{\alpha\beta} , \;
 [s , t^\dagger_{\alpha} ] = 0,
 \end{equation}
 we can reproduce the S=$\frac{1}{2}$, SU(2)
 algebra of the spins ${\bf S}_l$ and ${\bf S}_r$ :
 \begin{eqnarray}
 [S_{m \alpha}, S_{m \beta} ] =  {\it i}
{\epsilon}_{\alpha\beta\gamma}
 S_{m\gamma} , m=l,r ,\; [S_{l \alpha}, S_{r \beta} ] =0,
 \nonumber
 \end{eqnarray}
\begin{equation}
 \vec{S}_l\cdot\vec{S}_r = -\frac{3}{4} s^\dagger s +
\frac{1}{4} t^\dagger_
 \alpha t_\alpha, \;  S^2_l = S^2_r = \frac{3}{4}.
 \end{equation}

Substituting the operator representation of spins defined
in Eqs.(2.3) and (2.4) into the original
Hamiltonian of Eq.(2.1) we obtain the following form:
 \begin{equation}
 H = H_0 + \lambda (H_1 + H_2),
 \end{equation}
 where,
 \begin{eqnarray}
 H_0 &=& \sum_{i} J(-\frac{3}{4} s^\dagger_i s_i
 + \frac{1}{4} t^\dagger_{i\alpha} t_{i\alpha} ) - \sum_{i} \mu _i
  ( s^\dagger_i s_i + t^\dagger_{i\alpha}  t_{i\alpha} -1 ),
 \\
 H_1 &=&+ \frac{J}{2} \sum_{i} t^\dagger_{i\alpha}  t_{i+1\alpha}
 s^\dagger_{i+1} s_i +
t^\dagger_{i\alpha} t^\dagger_{i+1\alpha} s_i s_{i+1} + h.c.),
 \\
 H_2 &=&-\frac{J}{2} \sum_{i} \frac{1}{2}(1-\delta_{\alpha\beta})
 (t^\dagger_{i\alpha}t^\dagger_{i+1\alpha}t_{i\beta}t_{i+1\beta} -
 t^\dagger_{i\alpha}t^\dagger_{i+1\beta}t_{i+1\alpha}t_{i\beta}+h.c.).
 \end{eqnarray}
The part of the Hamiltonian  containing triple "t"
operators vanishes identically in the
present case due to reflection symmetry \cite {symm}. A
 site-dependent chemical potential $\mu _i$ is
introduced to impose the constraint of Eq.(2.6).
The Hamiltonian of Eq.(2.8) can now be solved by a
 mean-field decoupling of the quartic terms. This
yields an effective Hamiltonian $H_{mR}$ with only
quadratic operators. We take $<s_i> = \bar{s}$, which
means that the "s" bosons are condensed. We replace the
local constraint $\mu _i$ by a global one $\mu$ in accordance
 with the translational invariance of the problem along the
ladder-axis ($\hat{y}$-axis here). We will consider here only
 the terms $H_0$ and $H_1$ in Eq.(2.8) and will show later
 in the Appendix that inclusion of $H_2$ changes the results
only slightly. We perform a Fourier transformation of the
operators $
 t^\dagger_{i\alpha} = \frac{1}{\sqrt{N}}
\sum_{k} t^\dagger_{k\alpha}
  e^{ik r_i}$, where N is the number
of dimers or rungs in the ladder and k is the
wave-vector whose only non-zero component is along the
ladder axis. Thus, retaining only terms $H_0$ and $H_1$
in Eq.(2.8) we obtain the mean-field limit,
 \begin{equation}
 H_{m}(\mu,\bar{s}) = N(-\frac{3}{4}J{\bar{s}}^2
-\mu{\bar{s}}^2 +\mu ) +
\sum_{k} [ \Lambda_k t^\dagger_{k\alpha}t_{k\alpha}
+   \Delta_k(t^\dagger_{k\alpha}t^\dagger_{-k\alpha}
+ t_{k\alpha}t_{-k\alpha} )],
 \end{equation}
 where,
\begin{eqnarray}
\Lambda_k&=& \frac{J}{4} - \mu + \lambda J {\bar{s}}^2 \cos{k},
\nonumber
\\
\Delta_k&=& \frac{\lambda}{2} J {\bar{s}}^2 \cos{k}.
\end{eqnarray}
We have taken the lattice constant to be unity.
The ground state wave-function can be written in the form
$|\phi> = C exp[\sum_{i}\bar{s}_i s^\dagger_i
-\sum_{k} b_kt^\dagger_{k\alpha}t^\dagger_{-k\alpha}]|0>$.
The above mean-field Hamiltonian Eq.(2.12) can be
diagonalised by a Bogoliubov transformation
into new Boson operators $\gamma_{k\alpha}$ given by,
\begin{equation}
\gamma_{k\alpha} = \cosh {\theta_k} t_{k\alpha} +
\sinh {\theta_k}t^\dagger_{-k\alpha},
\end{equation}
where the coefficients  $\cosh {\theta_k}$ and
$\sinh {\theta_k}$ obtained in terms of
$\Lambda_k$, $\Delta_k$ and $\omega_k$ are given by,
\begin{eqnarray}
\cosh^{2}\theta_k &=&\frac{1}{2}
( \frac{\Lambda_k}{\omega_k} + 1), \nonumber
\\
\sinh^{2}\theta_k &=& \frac{1}{2} sgn(\Delta_k )
( \frac{\Lambda_k}{\omega_k} - 1).
\end{eqnarray}
We finally obtain,
\begin{equation}
H_{m}(\mu,\bar{s}) = N(-\frac{3}{4}J{\bar{s}}^2
-\mu{\bar{s}}^2 +\mu ) -
 \frac{N}{2} (\frac{J}{4} - \mu) + \sum_{k} \omega_k
( \gamma^\dagger_{k\alpha}\gamma_{k\alpha} +  \frac{1}{2} ),
\end{equation}
where,
\begin{equation}
\omega_k = [{\Lambda^2}_k - (2\Delta_k)^2 ]^{\frac{1}{2}}.
\end{equation}
The parameters $\mu$ and $\bar{s}$ are determined
 by solving the saddle point
equations;
\begin{equation}
 <\frac{\partial H_{m}}{\partial\mu} > = 0 ,
 <\frac{\partial H_{m}}{\partial\bar{s}} > = 0 .
\end{equation}

We obtain the following self-consistent eqations from
Eq.(2.18) evaluated at T=0.
\begin{eqnarray}
(\bar{s}^2-\frac{3}{2} ) + \frac{1}{2\pi} [ \sqrt{1+d}
{\bf E}(\sqrt{\frac{2d}{1+d}}) + \frac{1}{\sqrt{1+d}}
{\bf K}( \sqrt{\frac{2d}{1+d}} ) ] &=& 0, \nonumber
\\
(\frac{3}{2} + 2 \frac{\mu}{J} ) + \frac{2\lambda}{\pi d}
[ \frac{1}{\sqrt{1+d}} {\bf K}( \sqrt{\frac{2d}{1+d}} ) -
\sqrt{1+d} {\bf E}( \sqrt{\frac{2d}{1+d}} ) ] &=& 0.
\end{eqnarray}
where {\bf K}($\eta$) and {\bf E}($\eta$) are respectively
the complete elliptic integrals of the first and
second kind of modulus $\eta$. The dimensionless parameter
d is defined as
\begin{equation}
\displaystyle
d = \frac{2\lambda {\bar{s}}^2 }{(\frac{1}{4} - \frac{\mu}{J} )}.
\end{equation}

\section{RESULTS}
\subsection{Spin-triplet spectrum}

For each $\lambda$, the self-consistent solutions of
$\bar{s}$ and $\mu$ are
obtained from Eqs.(2.19) and are used to determine the excitation
spectrum of
the system of a single ladder obtained from Eq.(2.17) as,
\begin{equation}
\omega _k = J {(\frac{1}{4} - \frac{\mu}{J} )} [1 + d \cos{k}
]^{\frac{1}{2}}.
\end{equation}
These quasiparticle excitations arising from the spin-triplet
states of the spin-ladder form a band whose bandwidth
is a function of $\lambda$. The band-minimum is at k=$\pi$ and
in Fig. 2(b) we present plots of $\omega _k$/J
(relative to the band minima) as a function
of k for $\lambda$=0.1,0.5 and 1.0. We notice that the
dispersion around the band minimum gets more linear
with increasing $\lambda$ which is reminiscent of the case of a
linear chain ($\lambda = \infty$).

Recently extensive numerical calculations have been
performed \cite{barnes} on Heisenberg spin ladders using
Lanczos techniques. The spin-triplet dispersion
relations were obtained on a ladder with 2$\times$12 sites and
 for various values of $\lambda$. The results of such
a calculation are shown as filled circles
in Fig. 2(b) and the agreement with the present work is very good.
For very small values of $\lambda$
the spin-triplet excitation spectrum has a bandwidth
of 2$\lambda$J which is in excellent agreement with
those obtained from the Lanczos data and from the
strong coupling expansions of Heisenberg ladders \cite{barnes}.
The dispersion relation of Eq.(3.1) can be parametrized
by a spin-wave velocity which is given by
 $c_s=J(\frac{1}{4} - \frac{\mu}{J}) ({\frac{d}{2}})^{\frac{1}{2}}$.
The spin-wave velocity reduces to 0 as
$\lambda \rightarrow $0 and is 1.1J for $\lambda =1.0$.

\subsection{Spin Gap}

If the excitation spectrum $\omega _k$ is real and
positive everywhere in the Brillouin-zone then the
system is in a magnetically disordered (spin-liquid)
phase and has a spin gap given by,
\begin{equation}
\Delta = J {(\frac{1}{4} - \frac{\mu}{J} )} [1 - d]^ \frac{1}{2} .
\end{equation}
This is indeed found to be the case for a single
Heisenberg antiferromagnetic ladder.
In Fig. 2(c) we present a plot (continuous line) of the
spin gap (in units of J) as a function of
$\lambda$ obtained numerically by solving the non-linear
Eqs.(2.19) and then substituting the self-consistent
solutions $\bar{s}$ and $\mu$ into Eq.(3.2). Alternatively,
Eq(2.19) and Eq.(2.20) can be combined to give the following
single equation for d
\begin{equation}
d = \lambda [3.0 - \frac{2}{\pi} \frac{1}{\sqrt{1+d}} {\bf K}(
\sqrt{\frac{2d}{1+d}} )  ],
\end{equation}
which is easier to solve numerically. The solution of d for a
given $\lambda$ can be used in Eq.(2.19) to determine
$\mu$ and the spin gap is then obtained from Eq.(3.2)

We can analytically study the asymptotic behavior of the
 spin gap. For small values of $\lambda$ (and hence d) the
elliptic integrals of Eq.(2.19) can be
expanded in a power series as,
\begin{eqnarray}
E(\eta)&=& \frac{\pi}{2} (1 - \frac{1}{4} \eta^2 -
\frac{3}{64} {\eta}^4 - \cdots ), \nonumber
\\
K(\eta)&=& \frac{\pi}{2} (1 + \frac{1}{4} \eta^2 +
\frac{9}{64} {\eta}^4 - \cdots ),
\end{eqnarray}
and we obtain,
\begin{eqnarray}
d &=& 2 \lambda [ 1 - \frac{3}{8} {\lambda}^2 + O({{\lambda}^4})],
\\
(\frac{1}{4} - \frac{\mu}{J} ) &=&1 + \frac{1}{4} {\lambda}^2 +
O({{\lambda}^4}),
\end{eqnarray}
and the spin gap is given by
\begin{equation}
\Delta = J (1 - \lambda - \frac{{\lambda}^2}{4}  +
 \frac{{\lambda}^3}{8} + O({{\lambda}^4} )).
\end{equation}
This is consistent with Fig. 2(c) which shows a linear
drop of the spin gap $\Delta$ for
small values of $\lambda$. As $\lambda$ gets larger
the spin gap as shown in Fig. 2(c) deviates considerably
from the linear behavior and does not exhibit any
critical value for $\lambda$ where the gap vanishes.
Thus starting from the limit of $\lambda=0$ with
spin singlets on each rung of the ladder, we see that
any finite coupling $\lambda$ along the legs of the ladder
delocalizes the singlets on the rungs thereby reducing
the magnitude of the gap but not closing it completely.

The spin gap seems to approach zero as $\lambda
\rightarrow \infty$ which should be the right gapless limit
corresponding to decoupled spin-$\frac{1}{2}$ chains.
However, on closer inspection of Eq.(3.3) we notice that at
$\lambda = \infty$, d is given by the
solution of the equation
$ \frac{1}{\sqrt{1+d}} {\bf K}( \sqrt{\frac{2d}{1+d}})
 - \frac{3\pi}{2}= 0$.
This gives a value of d which is very close to 1.
In addition it can be shown
using Eq.(2.19) that as $\lambda\rightarrow$1,
$\frac{\mu}{J}$ diverges as $\lambda \ln (1-d)$. Hence the
spin gap given by Eq.(3.2) also
diverges as $\lambda \rightarrow \infty$. In the
present case the spin gaps start to increase for
values of $\lambda \geq$ 3 and the mean-field treatment
ceases to be valid.

The Heisenberg ladder has also been extensively studied
numerically \cite{dag},\cite{barnes} using the Lanczos
technique and the spin gaps have been determined on
2$\times$N ladders with N=4,6,8,10, and 12.
An extrapolation of these results to the bulk limit
is shown (as filled circles) in the inset of Fig. 2(c) as a
plot of $\frac{\Delta}{\lambda J}$ versus
${\lambda}^{-1}$ for values of $\lambda \leq 1$.
The full curve in the same inset indicates results
obtained from the present calculation. The agreement is
found to be good only for small values of $\lambda$.

The deviation of the present spin gaps from those
obtained numerically can be traced to the $\lambda^2$
terms in the expansion of the spin gap as
obtained in Eq.(3.7). In Ref.\cite{barnes} it was shown in
a strong coupling expansion that the spin gap varied
as $\Delta = J (1 -\lambda + \frac{3}{4} \lambda^2)$ for
small values of $\lambda$. The  $\lambda^2$ term arises
in the strong coupling expansion from short range effects
as the two nearest neighbor singlets surrounding the rung
which is excited to a triplet state are ineffective in
contributing to the energy of spin-triplet excitations to
$O(\lambda^2)$. Thus the positive co-efficient obtained for
the $\lambda^2$-term shifts the spin gaps above the 1-$\lambda$
line as $\lambda$ is increased in agreement with the numerical
results (see filled circles in the inset of Fig. (2c)).
However, in the present treatment a positive coefficient
is obtained in the expansion for $(\frac{1}{4} -\frac{\mu}{J})$
(Eq.3.6) but the expansion of (1-d)$^{\frac{1}{2}}$ produces an
overall negative co-efficient for the $\lambda^2$-terms which
shifts the spin gaps below the 1-$\lambda$ line as
$\lambda$ is increased. This implies that for a finite
 $\lambda$ the singlet and triplet levels are not pushed
apart enough due to the wrong treatment of the short-range
effect in the present mean-field method. This arises from
treating the local constraint to be valid only on the average.
We have therefore added an additional self energy term
of $O(\lambda^2)$ in the triplet levels of Eq.(2.9) which
corrects for the neglect of short range effect in the
present treatment.  We use the self enrgy term $\beta\lambda^2$
with an optimum value of $\beta=0.7$ which gives
reasonable values of the spin gaps for $\lambda \leq 1$ as
shown by the dashed line in the inset of Fig. 2(c). It should
be pointed out that this self energy correction does not
modify the dispersion curves of the spin-triplets shown in
Fig. 2(a), it merely shifts the positions of the minima at
$k=\pi$.

\subsection{Ground State Energy}

The ground state energy obtained from Eq.(2.16) is
given by (neglecting H$_2$ term)
\begin{equation}
\frac{E_G}{NJ} =  ( -\frac{3}{4} {\bar{s}}^2 -\tilde{\mu}{\bar{s}}^2
+\tilde{\mu} ) - \frac{1}{2} (\frac{1}{4} - \tilde{\mu} ) +
\frac{1}{\pi}(\frac{1}{4} - \tilde{\mu} )(1+d)^{\frac{1}{2}} {\bf E}(
\sqrt{\frac{2d}{1+d}} ),
\end{equation}
where d is defined in Eq.(2.20) and $\tilde{\mu} = \frac{\mu}{J}$.

In the limit of $\lambda \rightarrow 0$ the correct
ground state energy $\frac{E_G}{NJ} =-\frac{3}{4}$
is recovered. As $\lambda$ is switched on
the ground state energy decreases and we present in
Table I the values of $\frac{E_G}{2NJ}$ obtained
from Eq.(3.8) for certain values of $\lambda$.
As discusssed before for the spin gap the ground state
energies have also been obtained numerically
\cite{barnes} on finite length ladders and the
extrapolated values to the bulk limit is also
presented in Table I. We observe that the energies
obtained from the present mean-field treatment compare well
with numerical estimates for $\lambda \leq 0.5$.

\subsection{Structure Factor}

Structure factors are important physical quantities
which can be measured experimentally. The dynamic
spin-structure factor is defined as
\begin{equation}
S^{0,\pi} (q,\omega) =\int  dt e^{i\omega t}
< S^{0,\pi}_{q,z} (t) S^{0,\pi}_{-q,z}(0)>,
\end{equation}
where $S^{0,\pi}_{q,\alpha}$ represents the Fourier
transform of the $\alpha$-th component of the spin-operator
combination along the rungs given by
\begin{equation}
S^{0,\pi}_{q,\alpha} (t)  = \sum_{i} e^{i \vec{q} \cdot \vec{r_i}}
[ S_{l_{i,\alpha}} (t) \pm S_{r_{i,\alpha}} (t) ],
\end{equation}
where ${\bf S} _{l_{i}}$ and ${\bf S} _{r_{i}}$
are the two spin-operators at each
rung i of the ladder as shown in Fig. 2(a). Using the
representations of the spins ${\bf S} _l$ and ${\bf S} _r$
in terms of the singlet and triplet
operators as defined in Eqs.(2.3)and (2.4) we obtain
\begin{eqnarray}
 S_{l_{i,\alpha}} + S_{r_{i,\alpha}} &=& -
{\it i} {\epsilon}_{\alpha\beta\gamma}
t^\dagger_{i\beta} t_{i\gamma},  \nonumber
\\
S_{l_{i,\alpha}} - S_{r_{i,\alpha}} &=& s^\dagger_i t_{i\alpha}  +
t^\dagger_{i\alpha} s_i.
\end{eqnarray}
Substituting these in Eq.(3.10) and transforming the
"t" operators to Bogoliubov operators defined in Eq.(2.14)
we obtain the following expressions for the dynamic spin
structure factors;
\begin{eqnarray}
 {S^\pi}(q,\omega)  = \frac{1}{3}{\bar{s}}^2 (cosh2\theta_q
-sinh2\theta_q)(n_q + \Theta (\omega))\delta (\omega_q  - |\omega|),
\\
{S^0}(q,\omega) = \frac{1}{9} \sum_{k}\{
[cosh2(\theta_{k+q}-\theta_k)+1]
n_k(1+n_{k+q})\delta(\omega_{k+q}-\omega_k-\omega) \nonumber
\\
+\frac{1}{2}[cosh2(\theta_{k+q}-\theta_k)-1](n_k +
\Theta (\omega))(n_{k+q}
+ \Theta (\omega))\delta(\omega_{k+q}+\omega_k-|\omega|)\},
\end{eqnarray}
where,$\Theta (\omega)$ is a step function and
n$_k$ is the Bose occupation factor and $\cosh \theta_k$ and
$\sinh \theta_k$ are defined in Eq.(2.14)

The first term of Eq.(3.13) represents the simultaneous
emission and absorption of excitations and vanishes
identically at T=0 while the second term corresponds to
creation or annihilation of two excitations.
In Fig. 3 we present the results of the structure factors
$S^{0,\pi} (q,\omega)$ as a function of $\omega$ for
q=$\pi$ and for $\lambda=1.0$ after applying the
self-energy correction described at the
end of section III(b). The dominant contribution to the structure
factor comes from $S^\pi (q,\omega)$. The static structure factors
$S_{st}^{0,\pi} (q) =\int d\omega S^{0,\pi}(q,\omega)$
show peaks at the commensurate wavevector of q= $\pi$.
The spin-spin correlations in the present case decay
exponentially at large distances with a
correlation length given by,
\begin{equation}
\xi = \frac{c_s}{\Delta}=[{\frac{d}{2(1-d)}}]^{\frac{1}{2}},
\end{equation}
where $c_s$ is the spin-wave velocity and $\Delta$
the spin gap and d is as defined in Eq.(2.20). We notice that
$\xi \rightarrow$ 0 as $\lambda
\rightarrow$ 0 and for $\lambda=1$ we obtain a value of $\xi=2.66$.

\section{Double and periodic array of ladders}

We now consider the effects on the spin gap of
increasing the number of rungs in the
ladder. We first consider the case of a simply connected
double rung ladder shown in Fig. 4(a).
The two spin-ladders (denoted by left(L) and right(R))
are connected with a strength of $\lambda^\prime$J and
in each individual ladder we assume as before an
interaction strength of J along the rungs and $\lambda$J
along the legs of the ladders. The Hamiltonian of the system
can be written as,
\begin{equation}
H = \sum_{\ell} [(H_{o_\ell} + \lambda H_{1_\ell} ) + \frac{1}{2}
\lambda^\prime H^\prime_{1_\ell} ],
\end{equation}
where, $\ell$ denotes the ladder index, Left(L) or Right(R),  and
\begin{eqnarray}
H_{o_\ell} &=& J\sum_{i} (-\frac{3}{4} s^\dagger_{i_\ell}
s_{i_\ell} + \frac{1}{4} t^\dagger_{i_\ell\alpha}
 t_{i_\ell\alpha} )- \sum_{i \in R_\ell}
 {\mu}_{i} ( s^\dagger_{i_\ell} s_{i_\ell}+
t^\dagger_{i_\ell\alpha}  t_{i_\ell\alpha} -1 ), \nonumber
\\
H_{1_\ell} &=& \frac{J}{2} \sum_{i}
({t^\dagger}_{i_\ell\alpha}
t_{i+1_\ell\alpha} s^\dagger_{i+1_\ell} s_{i_\ell}+
t^\dagger_{i_\ell\alpha}
t^\dagger_{i+1_\ell\alpha} s_{i_\ell} s_{i+1_\ell} + h.c.),
\\
H^\prime_{1_\ell} &=& -\frac{J}{4} \sum_{i,{\ell\prime\neq\ell}}
(t^\dagger_{i_\ell\alpha}  t_{i_\ell\prime\alpha} s^\dagger_{i_\ell}
s_{i_\ell\prime}+ t^\dagger_{i_\ell\alpha}
t^\dagger_{i_\ell\prime\alpha}
s_{i_\ell} s_{i_\ell\prime} + h.c.). \nonumber
\end{eqnarray}

Here, we have neglected terms of the form H$_2$ of
Eq.(2.11) as it is shown in the appendix not
to change the results significantly in the parameter
range $0 \leq \lambda \leq 1$, which is of interest here.
It should be pointed out here that the
Hamiltonian containing triple "t" operators will
not vanish here in the H$^\prime$ part of the
Hamiltonian connecting the two ladders as it did in the
previous case along the ladder axis. However this
will not give any contribution in the magnetically
disordered phase when we take averages over the operators
in the mean-field decoupling.

Replacing $<s_{i_\ell}> = \bar{s}$ and $\mu_{i_\ell} = \mu$
in Eq.(4.2) and
taking Fourier transforms of the t-operators we obtain
 \begin{eqnarray}
 H_{m}(\mu,\bar{s}) = 2N(-\frac{3}{4}J{\bar{s}}^2
-\mu{\bar{s}}^2 +\mu ) +
\sum_{k,\ell} [ \Lambda_k t^\dagger_{k_\ell\alpha}t_{k_\ell\alpha}
 + \Delta_k
(t^\dagger_{k_\ell\alpha}t^\dagger_{-k_\ell\alpha} \nonumber
\\
+ t_{k_\ell\alpha}t_{-k_\ell\alpha} ) + \frac{\Gamma_k}{2}
\sum_{\ell\prime\neq\ell} (2 t^\dagger_{k_\ell\alpha}
t_{k_\ell\prime\alpha}
+t^\dagger_{k_\ell\alpha}t^\dagger_{-k_\ell\prime\alpha} +
t_{k_\ell\alpha}t_{-k_\ell\prime\alpha} )].
\end{eqnarray}
where $\Lambda_k$ and $\Delta_k$ are as defined in
Eq. (2.13) and $\Gamma_k = -\frac{\lambda^\prime}{4} {\bar{s}}^2J$.
We perform a Bogoliubov transformation
into two new Boson operators defined in terms of the
t-operators of the left and right hand ladders as
\begin{equation}
\gamma_{1,2_{k\alpha}} = \frac{1}{\sqrt{2}}
[(\cosh {\theta_k} t_{k_L\alpha} +
\sinh {\theta_k}t^\dagger_{-k_L\alpha})
\pm (\cosh {\theta_k} t_{k_R\alpha} +
\sinh {\theta_k}t^\dagger_{-k_R\alpha})].
\end{equation}
These are simply symmetric (bonding) and anti-symmetric
(anti-bonding) combinations of the individual
transformations in the left and right ladders.
We can now diagonalise the Hamiltonian of Eq.(4.3) using
this transformation and we obtain the following
\begin{equation}
H_{m}(\mu,\bar{s}) = 2N(-\frac{3}{4}J{\bar{s}}^2
-\mu{\bar{s}}^2 +\mu ) -
N(\frac{J}{4} - \mu) + \sum_{k,b=1,2}[ \omega_{b_k}
( \gamma^\dagger_{b_{k\alpha}}\gamma_{b_{k\alpha}}
+  \frac{1}{2} ),
\end{equation}
where $\omega_{1,2_k}$ is defined as,
\begin{equation}
\omega_{1,2_k}=[{\Lambda_k}^2 -(2\Delta_k)^2
\pm 2\Gamma_k (\Lambda_k -2\Delta_k )]^{\frac{1}{2}}.
\end{equation}
Thus the spin-triplet excitation spectrum of a double
ladder with three rungs and four legs consists of
two branches for each $\lambda$ and $\lambda^\prime$
representing the bonding and the anti-bonding states.
The splitting of these two branches is governed by the
transfer matrix proportional to $\Gamma_k$ which is in turn
proportional to $\lambda^\prime$. As $\lambda^\prime
\rightarrow 0$ the two branches collapse into a single
branch of the one rung ladder described in section III(a).
We have plotted in Fig. 4(b) the
spin-triplet excitation spectrum for the double
ladder with the value $\lambda = \lambda^\prime =1$
and after applying the self-energy correction in each
ladder as described at the end of section III(b). For
comparison we show in the same figure a plot
(shown as the dashed curve) of the
excitation spectrum of a single rung ladder
($\lambda^\prime =0$).

Following the same procedure as described in section
II we can write down the
mean-field equations evaluated at T=0 as
\begin{eqnarray}
(\bar{s}^2-\frac{3}{2} ) &=&- \frac{1}{4}\int
\frac{dk_y}{2\pi}
(\frac{1+\frac{d}{2}a_+}{[1+da_+]^\frac{1}{2}}
 + \frac{1+\frac{d}{2}a_-}{[1+da_-]^\frac{1}{2}}),  \nonumber
\\
(\frac{3}{2} + 2 \frac{\mu}{J} ) &=&
\frac{\lambda}{2} \int\frac{dk_y}{2\pi}
(\frac{a_+}{[1+da_+]^\frac{1}{2}} +
\frac{a_-}{[1+da_-]^\frac{1}{2}}),
\end{eqnarray}
where d is defined in Eq.(2.20), and $a_\pm$ is given by
\begin{equation}
a_\pm = \cos{k} \pm \frac{\lambda^\prime}{4\lambda}.
\end{equation}
The expression for the spin gap is
\begin{equation}
\Delta = J {(\frac{1}{4} - \frac{\mu}{J} )} [1 -
d(1+\frac{\lambda^\prime}{4\lambda})]^ \frac{1}{2}.
\end{equation}
We have plotted in Fig. 4(c) the spin-gaps of the double
ladder as a function of $\lambda^\prime$ for different values
of $\lambda$. We obtain a value of 0.1J for
the spin gap for $\lambda=\lambda^\prime=1$ and
we notice that for each $\lambda$ the
spin gap reduces drastically with $\lambda^\prime$.
This can be explained by the fact the singlets along
the rungs of the two ladders can not only delocalize
along the ladder axes via the $\lambda$-coupling but also
across the ladders through the transfer matrix proportional
to $\lambda^\prime$. This produces for any non-zero
coupling of two single rung ladders ($\lambda^\prime \ne 0$)
a splitting in the excitation spectrum of the individual
ladders into two branches, one above (the antibonding branch)
and the other below (the bonding branch) as shown in Fig. 4(b).
Since the spin gap is a measure of the minimum energy
in the excitation spectrum (at $k=\pi$) and since one of the
branches of the double ladder is always lower than that of
a single ladder for any non-zero $\lambda^\prime$ we can
conclude that the spin gap of a double ladder is lower
than that of single ladder and the spin gap should
progressively decrease on increasing the width of the ladder.

In the same context it is worthwhile to consider
a periodic array of ladders,with $\lambda$=1 in each
ladder, and simply connected to one
another by $\lambda^\prime$ and then to ask the question
for what value of $\lambda^\prime$ should the
spin gap disappear since it is
known that the ground state of a two-dimensional
antiferromagnet has longrange
order. We can study the system of ladders arranged
periodically within the
formalism described previously and the self-consistent
equations are given by
\begin{eqnarray}
(\bar{s}^2-\frac{3}{2} ) &=&- \frac{1}{2}\int\!\!\int
\frac{d^2k}{(2\pi)^2}
\frac{1+\frac{d}{2}a}{[1+da]^\frac{1}{2}},\nonumber
\\
(\frac{3}{2} + 2 \frac{\mu}{J} ) &=&+\lambda
\int\!\!\int\frac{d^2k}{(2\pi)^2} \frac{a}{[1+da]^\frac{1}{2}}.
\end{eqnarray}
where {\bf k} is now a two-dimensional wave-vector
with components $k_y$ along the ladder-axis and
$k_x$ across the ladders.
The parameter d is define in Eq.(2.20) and $a$ is given by
\begin{equation}
a = \cos{k_y} - \frac{\lambda^\prime}{2\lambda}\cos{k_x}.
\end{equation}
Combining the equations  in Eq.(4.10) we can write down
the following  single mean field equation for d which
is easier to solve numerically.
\begin{equation}
\frac{d}{\lambda} =3.0-
\int\!\!\int\frac{d^2k}{(2\pi)^2} \frac{1}{[1+da]^\frac{1}{2}}.
\end{equation}

The excitation spectrum of the periodic system of
ladders has a minimum at ${\bf k} =(\pi,\pi)$ and the
the spin gap given by
$\Delta = J {(\frac{1}{4} - \frac{\mu}{J} )}
[1-d(1+\frac{\lambda^\prime}{2\lambda})]^\frac{1}{2}$
is plotted in Fig. 5 as a
function of $\lambda^\prime$ with $\lambda$ set to 1.0.
We have again applied the self-energy corrections on
each individual ladders as described at the end of
section III(b). We notice that the spin gap vanishes
for a value of $\lambda^\prime \approx 0.25$.
As described before for the double ladder
the decrease in the spin gap with $\lambda^\prime$
is explained by the delocalization of the singlets
across the ladders. In this case the transfer
matrix connecting the ladders is twice as large
as in the case of a double ladder thereby
decreasing the spin gap with $\lambda^\prime$ even
faster than that of a double ladder.

\section{Frustrated Double ladders and Trellis lattice}

We now consider the double ladder described in the
previous section but connected to each other
through zig-zag frustrated couplings
$J^\prime = \lambda ^\prime J$ as shown in Fig.1(b),
where J is the standard rung coupling.
We retain the Hamiltonian in the
form of Eq.(4.2) but with H$^\prime_{1\ell}$
containing an additional term given by
\begin{equation}
H^\prime\prime_{1_\ell} =
-\frac{J}{4} \sum_{i,\ell\prime\neq\ell}
(t^\dagger_{i_\ell\alpha}
t_{i+1_\ell\prime\alpha} s^\dagger_{i_\ell}
s_{i+1_\ell\prime}+ t^\dagger_{i_\ell\alpha}
t^\dagger_{i+1_\ell\prime\alpha}
s_{i_\ell} s_{i+1_\ell\prime} + h.c.),
\end{equation}

The mean field Hamiltonian and the excitation
spectrum are those described by Eqs.(4.5) and (4.6) with
$\Gamma_k$ replaced by a new form $\Gamma_k =
-\frac{\lambda^\prime}{4}\bar{s}^2J(1+\cos{k})$
We obtain the self-consistent equations of Eq.(4.7)
but with a$_\pm$ given by
$a_\pm = \cos{k} \pm \frac{\lambda^\prime}{4\lambda}(1+cos{k})$

The excitation spectrum of the double ladder with a zig-zig
frustrated coupling also consists of two branches like in the
case of a simply connected double ladder. However, in the present
case at the minimum position of the spectrum ($k=\pi$) the
two branches become degenerate. This can be explained by
the fact that the singlets on two successive rungs of each
ladder are completely out of phase and hence are not able to
delocalise across the ladders through the zig-zag frustrated
couplings $\lambda^\prime$. Thus the spin gap
defined by the minimum value of the spectrum at $k=\pi$ is now
given by,
\begin{equation}
\Delta = J {(\frac{1}{4} - \frac{\mu}{J} )} [1-d]^\frac{1}{2},
\end{equation}
which does not contain the $\lambda^\prime$ term explicitly.
Thus the spin gaps should not change much form the values
obtained for a single ladder and this is indeed what is observed
as shown by the dashed lines in Fig. 4(c). We have again applied
the self-energy correction to each individual ladder. We
notice from Fig. 4(c) that for each $\lambda$ the spin gap
slightly increases with $\lambda^\prime$. This
simply reflects the slight changes in the self-consistent
solutions of d and $\mu$ with $\lambda^\prime$.

We can extend the results discussed above to the
trellis lattice shown in Fig. 1(b). The excitation spectrum
would now be a function of {\bf k} with components $k_y$ along
the ladder axis and $k_x$ across the ladders.
At the point $k_y=\pi$ the spectrum would be
dispersionless with $k_x$ as a result of destructive interference
of successive rung-singlets along the ladder axis.
Thus a frustrated coupling between the ladders in the trellis
lattice will not affect the spin gaps of each individual ladder
and hence it will only help in retaining the spin-liquid
nature.

\section{Conclusions}

We have shown that the CuO double chains intergrown periodically
in the CuO$_2$ planes of the new series of infinite-layer
compounds Sr$_{n-1}$Cu$_{n+1}$O$_{2n}$ (n=3,5,7,...)
creates a new two-dimensional spin-lattice, the trellis lattice.
Such a lattice can be described by Heisenberg spin ladders with
$n_r=\frac{1}{2}(n-1)$ rungs and $n_l=\frac{1}{2}(n+1)$ legs
with the usual antiferromagnetic coupling J inside each
ladder and a weak and frustrated interladder coupling J$^\prime \ll$J.
On neglecting J$^\prime$ the Cu-O planes
can be thought of as built up of independent
quasi one-dimensional ladders with odd n$_r$ and even n$_l$
(corresponding to n=3,7,11,...) and those with even
n$_r$ and odd n$_l$(corresponding to n=5,9,13,...) which exhibit
different spin excitation spectra. The
former are gapped with shortrange exponentially
decaying magnetic correlation functions
while the latter are gapless with a longrange
powerlaw decay of the correlation functions.

We have used the bond operator representation of
quantum spins in a mean field treatment with self-energy
correction to first study the excitations of the
simplest single rung ladder with two legs corresponding to n=3.
We have obtained the spin-triplet dispersion
with a minimum at $k=\pi$ and a spin-gap of
$(\approx \frac{1}{2} J)$. These are in good agreement
with numerical estimates \cite{dag},\cite{barnes}. It should be
pointed out that conventional spin-wave calculations applied
to a ladder \cite{jap},\cite{barnes} predict
gapless excitations for all values of the couplings
$\lambda J$ along the legs of the ladder,
even for $\lambda$=0 which is clearly unphysical.

We have been able to extend the mean field treatment
to double ladders and to a periodic array of ladders.
We find that increasing the rungs of the
ladder drastically reduces the spin gap.
For a double ladder with four legs and three rungs corresponding
to n=7 we obtain a spin-gap of only 0.1J. In a periodic
array of ladders with intraladder coupling J we find that the
spin gap vanishes for an interladder coupling of $\approx 0.25J$.

We have also studied the effect of a frustrated
coupling, such as that of a trellis lattice, introduced between
two ladders. We find a slight enhancement in the spin gap.
Extending the results to a trellis lattice we can
show that the spin-liquid nature will be preserved. Thus
stoichiometric Sr$_{n-1}$Cu$_{n+1}$O$_{2n}$
compounds with n=3,7,11 ... will be frustrated
quantum antiferromagnets  with a quantum disordered or
spin-liquid ground state.
The resulting trellis lattice will be a realization
of the short range RVB ground state for a
spin-$\frac{1}{2}$ system \cite{anderson2}.

The implications of these results to the other high
T$_c$ compound can be conjectured as follows.
Certain underdoped high T$_c$ samples have been
experimentally shown to have spin gaps and a
theoretical description of this
in terms of frustrated next nearest neigbour
coupling in a two-dimensional antiferromagnetic
lattice is obtained only with large and unphysical values of
the coupling. However if the two dimensional
Cu-O planes were thought of having some microstructure
(introduced upon doping)
then the spin gaps could be explained quite naturally.
\vspace{1cm}

\acknowledgements
We wish to thank R. Hlubina and H.Tsunetsugu for stimulating
conversations. This work was supported by grants
from the Swiss National Fonds.

\appendix
\section*{}
The results presented in section III were obtained by
neglecting the term H$_2$ of Eq. (2.11) containing four
triplet operators. Here we will study the changes
on including this term in a similar mean field treatment
as presented before. By taking quadratic decouplings of
the operators in H$_2$ and
performing Fourier transformations we obtain,
 \begin{equation}
 H_2 = -\frac{\lambda J}{3}[ \sum_{k} cosk_y [ 2P
t^\dagger_{k\alpha}t_{k\alpha} - Q (t^\dagger_{k\alpha}t^\dagger_
{-k\alpha} + t_{k\alpha}t_{-k\alpha} ) ] - N(P^2 -Q^2)] ,
 \end{equation}
where $\lambda$ and N are as defined previously and P and
Q are two new mean-fields defined as
\begin{eqnarray}
P &=& <t^\dagger_{i\alpha}t_{i+1\alpha}>, \nonumber
\\
Q &=& <t_{i\alpha}t_{i+1\alpha}>.
\end{eqnarray}

On including Eq.(A1) into Eq.(2.11) we obtain a mean
field Hamiltonian similar to Eq.(2.12)
\begin{eqnarray}
 H_{mR}(\mu,\bar{s},P,Q) = N(-\frac{3}{4}J{\bar{s}}^2
-\mu{\bar{s}}^2 +\mu )
 - \frac{N\lambda J}{3} (P^2 -Q^2)] \nonumber
\\
+ \sum_{k} [ \Lambda_k t^\dagger_{k\alpha}t_{k\alpha} +
\Delta_k (t^\dagger_{k\alpha}t^\dagger_{-k\alpha}
+ t_{k\alpha}t_{-k\alpha} )] ,
\end{eqnarray}
but with $\Lambda_k$ and $\Delta _k$ replaced by
\begin{eqnarray}
\Lambda_k&=& \frac{J}{4} - \mu + \lambda J {\bar{s}}^2 \cos{k}
- \frac{2\lambda}{3} PJ\cos{k} ,\nonumber
\\
\Delta_k&=&+\frac{\lambda}{2} J {\bar{s}}^2 \cos{k}
+\frac{\lambda}{3} QJ \cos{k} .
\end{eqnarray}
We can diagonalise Eq.(A3) by a Bogoliubov transformation
as described earlier and the spectrum is given by Eq.(2.17)
but with $\Delta _k$ and $\Lambda _k$
replaced by their new forms. The parameters $\mu$, $\bar{s}$,
P and Q are obtained by solving the saddle-point
equations which reduce at T=0 to the following equations;
\begin{eqnarray}
(\bar{s}^2-\frac{3}{2} ) &=& -\int \frac{dk}{2\pi}
\frac{\Lambda_k}{2\omega_k}, \nonumber
\\
(\frac{3}{2} + 2 \frac{\mu}{J} ) &=&\lambda \int \frac{dk}{2\pi}
\frac{\Lambda_k - 2 \Delta_k}{\omega_k} \cos{k}  ,\nonumber
\\
P&=&-\int \frac{dk}{2\pi} \frac{\Lambda_k}{2\omega_k}\cos{k}  ,
\\
Q&=&\int \frac{dk}{2\pi} \frac{\Delta_k}{\omega_k}
\cos{k}.\nonumber
\end{eqnarray}
By numerically solving for the mean-field parameters
the spin gap is obtained and is plotted in Fig. 2(b)
(dashed lines) as a function of $\lambda$. We
notice that the inclusion of $H_2$ does not change the
results significantly even for  $\lambda \approx 1$

\begin{figure}
\caption{(a) Schematic diagram of a single
copper-oxide sheet in Sr$_{n-1}$Cu$_{n+1}$O$_{2n}$
showing the parallel lines of CuO double chains.
The Cu-atoms are shown as big black dots while the
oxygen atoms are located at all the points of intersections
of the straight lines. The dashed regions correspond to the
usual square co-ordinated CuO$_2$ regions.
(b) The two-dimensional trellis lattice formed
from the exchange couplings in a single copper-oxide plane
of Sr$_{n-1}$Cu$_{n+1}$O$_{2n}$ at stoichiometry. The usual
antiferromagnetic coupling J describes each of the
ladders having n$_r =\frac{1}{2}$(n-1) rungs and
n$_l=\frac{1}{2}$(n+1) legs and the ladders are
coupled to one another through weak and frustrated
zig-zag couplings J$^\prime$.}
\label{fig1}
\end{figure}

\begin{figure}
\caption{(a) A single S=$\frac{1}{2}$ Heisenberg
antiferromagnetic ladder with couplings J along the rungs (i)
and $\lambda$J along the left(l) and right(r) legs of the ladder.
(b) The dispersions of the spin-triplet excited
states of the ladder of (a) relative to the band
minimum at k=$\pi$ for several values of $\lambda$.
The continuous curves are obtained from the present
mean-field method and the filled circles are the Lanczos
results of Ref.12 obtained on a $ 2\times 12$ ladder.
(c) The spin gap $\Delta$ (in units of J) as a
function of $\lambda$ obtained from the present
mean-field treatment (continuous curve). The dashed
curve is a similar plot after including higher order
terms as described in the Appendix. In the inset a plot of
$\Delta /\lambda J$ versus $\lambda^{-1}$ is shown.
Continuous curve is from the present treatment
(without higher order terms) and the filled circles are
the numerical results of Ref.8. The dashed line in the
inset is obtained after including
a self-energy correction [described at the end of
section III(b)] to the present treatment.}
\label{fig 2}
\end{figure}

\begin{figure}
\caption{The dynamical structure factors $S^{0,\pi}(q,\omega)$
as a function of frequency $\omega$ for q=$\pi$.}
\label{fig 3}
\end{figure}

\begin{figure}
\caption{(a) A double Heisenberg ladder consisting
of two single S=$\frac{1}{2}$ antiferromagnetic
ladders of Fig. 2(a) connected to each other
through the couplings $\lambda ^{\prime} J$.
(b) The dispersions of the spin-triplet excited
 states (bonding and anti-bonding states of Eq.(4.6))
of the double ladder of (a) for
$\lambda =\lambda ^\prime =1.0$. The dashed curve
is a similar plot for a single ladder ($\lambda^\prime=0$)
with $\lambda=1$. (c) The spin gap $\Delta$
(in units of J) as a function of $\lambda ^\prime$
for several values of $\lambda$. The dashed curves are
similar plots obtained for the case of a
double ladder connected to each other through
the zig-zag frustrated
couplings $J ^\prime =\lambda ^\prime J$  of Fig. 1(b).}
\label{fig 4}
\end{figure}

\begin{figure}
\caption{The spin gap (in units of the rung coupling J)
of a periodic arrangement of ladders as a function
of the interladder coupling $\lambda ^\prime$ and with
an intraladder couling of $\lambda=1$. }
\label{fig 5}
\end{figure}

\begin{table}
\caption{Results of the ground state energy
$\frac{E_G}{2NJ}$ of a Heisenberg ladder}
\begin{tabular}{lcr}
$\lambda$ &\multicolumn{2}{c}{$\frac{E_G}{2NJ}$} \\ \cline{2-3}
&Present work	&Ref.\cite{barnes}\\ \tableline
0.5&-0.394&-0.43\\
1.0&-0.475&-0.578\\ \tableline
\end{tabular}
\end{table}


\begin{thebibliography}{99}
\bibitem{symm} The ladder possesses a reflection plane
perpendicular to the rungs and passing through the centers
of the rungs. The terms containing the triple "t" operators
change sign under the reflection and hence vanish in the
evaluation of Eq.(2.8).
\end{thebibliography}
\end{document}